\documentclass[12pt,oneside,a4paper]{article}

\usepackage{hyperref}
\hypersetup{colorlinks,bookmarksopen,bookmarksnumbered,citecolor=blue,
linkcolor=black,pdfstartview=FitH,urlcolor=blue}

\usepackage{graphicx}
\usepackage{amssymb}
\usepackage{cite}
\usepackage{bm}
\usepackage{indentfirst}
\usepackage{amsmath}
\usepackage{hhline}
\usepackage{multirow}
\usepackage{color}

\allowdisplaybreaks
\numberwithin{equation}{section}

\definecolor{dred}{rgb}{0.7,0.15,0.09}
\definecolor{dblue}{rgb}{0,0.12,0.64}
\definecolor{dgreen}{rgb}{0.2,0.51,0.19}

\begin{document}

\begin{titlepage}

\begin{flushright}
KANAZAWA-22-01
\end{flushright}

\begin{center}

\vspace{1cm}
{\large\textbf{
Semi-annihilating dark matter coupled with Majorons
}
 }
\vspace{1cm}

\renewcommand{\thefootnote}{\fnsymbol{footnote}}
Takumi Miyagi$^{1}$\footnote[1]{t\_miyagi@hep.s.kanazawa-u.ac.jp}
and 
Takashi Toma$^{1,2}$\footnote[2]{toma@staff.kanazawa-u.ac.jp}
\vspace{5mm}

\textit{
 $^1${Institute for Theoretical Physics, Kanazawa University, Kanazawa 920-1192, Japan}\\
 $^2${\mbox{Institute of Liberal Arts and Science, Kanazawa University, Kanazawa 920-1192, Japan}}
}

\vspace{8mm}

\abstract{
The thermal production mechanism of dark matter is attractive and well-motivated by predictivity. 
A representative of this type of dark matter candidate is the canonical, weakly interacting massive particles. 
An alternative is semi-annihilating dark matter, which exhibits different phenomenological aspects from the former example. 
In this study, we constructed a model of dark matter semi-annihilating into a pair of anti-dark matter and a Majoron based on a global $U(1)_{B-L}$ symmetry, and 
show that semi-annihilation induces the core formation of dark matter halos, which can alleviate the so-called small-scale problems. 
In addition, the box-shaped spectrum of neutrinos was produced by the subsequent decay of the Majoron. 
This can be a distinctive signature of the dark matter in the model. 
We find a parameter space where the produced neutrinos can be detected by the future large-volume neutrino detector Hyper-Kamiokande. 
We also compared the dark matter scenario with the case of halo core formation by the strongly self-interacting dark matter. 
}

\end{center}
\end{titlepage}

\renewcommand{\thefootnote}{\arabic{footnote}}
\newcommand{\bhline}[1]{\noalign{\hrule height #1}}
\newcommand{\bvline}[1]{\vrule width #1}

\setcounter{footnote}{0}

\setcounter{page}{1}

\section{Introduction}

Weakly interacting massive particles (WIMPs) are one of the prominent thermal dark matter candidates and are being searched for by various experiments and observations such as 
indirect detection, direct detection, and collider production. 
However, no clear signal of WIMPs has been found so far, and the resultant bounds on the interactions between WIMPs and normal matter are getting stronger. 
In particular, recent direct detection experiments impose a stringent limit on the elastic scattering cross section with nuclei/nucleons. 
The XENON1T and PandaX-4T Collaborations set a limit of $4.1\times10^{-47}~\mathrm{cm}^2$ at a dark matter mass of $30~\mathrm{GeV}$~\cite{XENON:2018voc} 
and $3.8\times10^{-47}~\mathrm{cm}^2$ at $40~\mathrm{GeV}$~\cite{PandaX-4T:2021bab}, respectively.

Even in such a situation, thermally produced WIMPs are still attractive and well-motivated 
because the dark matter relic abundance can be determined independently from what occurred in the early universe. 
Furthermore, the correctness of the thermal history of the universe has been partially proven by successful Big Bang Nucleosynthesis (BBN). 
One of the ideas by which these strong limits are naturally avoided is to consider a velocity-dependent cross section such as pseudo-Nambu--Goldstone dark matter~\cite{Gross:2017dan} or 
pseudo-scalar interacting fermionic dark matter~\cite{Freytsis:2010ne}. 
This type of dark matter may be detected through experiments and observations if the dark matter is boosted by some mechanism, because the interaction itself is not very small.

Another method to avoid the strong constraint of direct detection experiments is to make dark matter one step closer to the dark sector, namely semi-annihilating dark matter where semi-annihilations include (anti-)dark matter particles in the final state~\cite{Hambye:2008bq, DEramo:2010keq}. 
The relic abundance of semi-annihilating dark matter is also thermally produced via the freeze-out mechanism, similar to WIMPs, whereas it exhibits some features different from the canonical WIMPs. 
For example, semi-annihilating dark matter may improve small-scale problems, such as core-cusp, too-big-to-fail, and missing satellite problems~\cite{Tulin:2017ara}, via the core formation of dark matter halos~\cite{Chu:2018nki, Kamada:2019wjo}.
The self-heating of dark matter by semi-annihilations also affects the calculation of the thermal relic abundance~\cite{Kamada:2017gfc}, 
and the evolution of cosmological perturbations~\cite{Kamada:2018hte}. 
In addition, semi-annihilations produce (semi-)relativistic boosted dark matter in galaxies and exhibit distinctive signatures not induced by canonical WIMPs~\cite{Agashe:2014yua, Toma:2021vlw}. 

On the other hand, the mechanism of small neutrino mass generation is still unknown, although neutrino oscillation experiments have confirmed that neutrinos are massive~\cite{deSalas:2020pgw}. 
The canonical seesaw mechanism is the simplest possibility for small neutrino mass generation~\cite{Minkowski:1977sc, Yanagida:1979as, Gell-Mann:1979vob}; 
however, the mass scale of the right-handed neutrinos is undetermined. 
It may be correlated with a spontaneous breaking of global or gauge symmetries. 

In this study, we construct a model of semi-annihilating dark matter and simultaneously generate neutrino masses based on a global $U(1)_{B-L}$ symmetry. 
The Nambu--Goldstone boson associated with the global $U(1)_{B-L}$ symmetry, the so-called Majoron, can naturally be light enough compared with the other particles because of its nature. 
Thus, the semi-annihilation $\chi\chi\to\overline{\chi}J$ can easily be dominant over the other annihilation channels, where $\chi$ is the dark matter and $J$ is the Majoron.
We calculate the thermal relic abundance of dark matter and find the parameter space that can improve the small-scale problems via the self-heating effect by semi-annihilation 
taking into account the constraints of the BBN and perturbative unitarity of the couplings. 
Furthermore, we show that neutrinos are produced by the subsequent decay of the Majoron produced by the semi-annihilation $\chi\chi\to\overline{\chi}J\to\overline{\chi}\nu\nu$, 
and these neutrinos indicate a characteristic box-shaped spectrum that can be detected in future neutrino large detectors such as Hyper-Kamiokande (HK)~\cite{Abe:2011ts, Bell:2020rkw}. 
Finally, for comparison, we also investigate the case in which the Majoron mass is larger than the dark matter mass. 
In this case, the three-to-two dark matter annihilations, such as $\chi\chi\chi\to\chi\overline{\chi}$, determine the relic abundance of dark matter 
instead of the semi-annihilation $\chi\chi\to\overline{\chi}J$.

\section{The model}

\begin{table}[t]
\begin{center}
\begin{tabular}{|c||c|c|c|c|c|c||c|c|c|}\hline
 & $Q$ & $u^c$ & $d^c$ & $L$ & $e^c$ & $H$ & $\nu_R^c$ & $\Phi$ & $\chi$\\
\hhline{|=#=|=|=|=|=|=#=|=|=|}
$SU(3)_c$ & $\bm{3}$ & $\overline{\bm{3}}$ & $\overline{\bm{3}}$ & $\bm{1}$ & $\bm{1}$ & $\bm{1}$ & $\bm{1}$ & $\bm{1}$ & $\bm{1}$\\\hline
$SU(2)_L$ & $\bm{2}$ & $\bm{1}$ & $\bm{1}$ & $\bm{2}$ & $\bm{1}$ & $\bm{2}$ & $\bm{1}$ & $\bm{1}$ & $\bm{1}$\\\hline
$U(1)_Y$ & $1/6$ & $2/3$ & $-1/3$ & $-1/2$ & $1$ & $1/2$ & $0$ & $0$ & $0$\\\hline
$U(1)_{B-L}$ & $1/3$ & $-1/3$ & $-1/3$ & $-1$ & $1$ & $0$ & $1$ & $2$ & $2/3$\\\hline
\end{tabular}
\caption{Particle contents and quantum charges of the model.}
\label{tab:1}
\end{center}
\end{table}

We propose a model of semi-annihilating dark matter that is dominantly coupled with neutrinos via a Majoron. 
The Standard Model (SM) is extended with a global $U(1)_{B-L}$ symmetry, and the particle contents and charge assignments are listed in Table ~\ref{tab:1}. 
The new particles other than the SM fields are three right-handed neutrinos $\nu_{Ri}^c~(i=1,2,3)$, the singlet complex scalar $\Phi=(\phi+iJ)/\sqrt{2}$ including a Majoron field $J$, and 
another singlet complex scalar $\chi$ which will be identified as dark matter.
The Lagrangian of the new particles invariant under the symmetry is given by
\begin{align}
 \mathcal{L}=-\left(y_{\nu}\tilde{H}\overline{L}\nu_R + \frac{y_\Phi}{2}\Phi\overline{\nu_R^c}\nu_R + \mathrm{h.c.}\right)-\mathcal{V}(H,\Phi,\chi),
\end{align}
where the generation indices are omitted. 
The scalar potential $\mathcal{V}(H,\Phi,\chi)$ is given by
\begin{align}
 \mathcal{V}(H,\Phi,\chi)=&
-\frac{\mu_H^2}{2}|H|^2-\frac{\mu_\Phi^2}{2}|\Phi|^2+\frac{\mu_\chi^2}{2}|\chi|^2
+\frac{\lambda_H}{2}|H|^4+\frac{\lambda_\Phi}{2}|\Phi|^4+\frac{\lambda_\chi}{2}|\chi|^4\nonumber\\
&
+\lambda_{H\Phi}|H|^2|\Phi|^2+\lambda_{H\chi}|H|^2|\chi|^2+\lambda_{\Phi\chi}|\Phi|^2|\chi|^2\nonumber\\
&
+\left(\frac{\lambda}{3\sqrt{2}}\Phi^*\chi^3-\frac{m_J^2}{4}\Phi^2+\mathrm{h.c.}\right).
\end{align}
The last term $m_J^2\Phi^2/4$ is the explicit global $U(1)_{B-L}$ breaking term, which is essential for giving a mass to the Nambu--Goldstone boson (Majoron), as we show in the following.
The CP phases in the couplings $\lambda$ and $m_J^2$ can be absorbed by the field redefinitions of $\chi$ and $\Phi$, respectively. 
Thus, all couplings in the scalar potential can generally be regarded as real parameters. 

\subsection{The scalar sector}
The electroweak gauge symmetry and global $U(1)_{B-L}$ symmetry are spontaneously broken by the vacuum expectation values (VEVs) of $H$ and $\Phi$, respectively. 
Then the fields $H$ and $\Phi$ can be parametrized as
\begin{align}
H=\frac{1}{\sqrt{2}}\left(
\begin{array}{c}
 0\\
 v+h
\end{array}
\right),\quad
\Phi=\frac{1}{\sqrt{2}}\left(v_\phi+\phi+iJ\right),
\end{align}
where $v$ and $v_\phi$ are the VEVs. 
The CP-even components $h$ and $\phi$ mix with each other, whereas the CP-odd component $J$ itself is the mass eigenstate 
being the Nambu--Goldstone boson associated with the global $U(1)_{B-L}$ symmetry, the so-called Majoron. 
Using the stationary conditions given by
\begin{align}
\mu_H^2&=\lambda_Hv^2+\lambda_{H\Phi}v_\phi^2,\\
\mu_{\Phi}^2&=\lambda_{\Phi}v_\phi^2+\lambda_{H\Phi}v^2-m_J^2,
\end{align}
the mass matrix of the CP-even components is written as
\begin{align}
\mathcal{V}\supset\frac{1}{2}\left(
\begin{array}{cc}
 h & \phi
\end{array}
\right)\left(
\begin{array}{cc}
 \lambda_Hv^2 & \lambda_{H\Phi}vv_\phi\\
\lambda_{H\Phi}vv_\phi & \lambda_\Phi v_\phi^2
\end{array}
\right)\left(
\begin{array}{c}
 h\\
\phi
\end{array}
\right).
\end{align}
This mass matrix can be diagonalized by a unitary matrix. 
As a result, the gauge eigenstates $h$ and $\phi$ are rewritten by the unitary matrix and mass eigenstates $h_1$ and $h_2$ as
\begin{align}
\left(
\begin{array}{c}
 h\\
\phi
\end{array}
\right)=\left(
\begin{array}{cc}
 \cos\theta & \sin\theta\\
-\sin\theta & \cos\theta
\end{array}
\right)\left(
\begin{array}{c}
 h_1\\
 h_2
\end{array}
\right),
\end{align}
where $h_1$ is the SM-like Higgs boson with a mass of $m_{h_1}=125~\mathrm{GeV}$, and $h_2$ is the additional Higgs boson with mass $m_{h_2}$. 

The Higgs mixing angle $\sin\theta$ is constrained by K and B meson decays, BBN and supernova explosions, which are summarized in Ref.~\cite{Winkler:2018qyg}. 
The upper bound is roughly given by $\sin\theta\lesssim 3\times10^{-4}$ for $m_{h_2}\lesssim 5~\mathrm{GeV}$ via K and B meson decays. 
For $m_{h_2}\lesssim 0.2~\mathrm{GeV}$, the lower bound $\sin\theta\gtrsim 10^{-5}$ is also imposed by BBN and supernova explosions. 
The Higgs invisible decays $h_1\to \chi\chi,h_2h_2$ also give an upper bound on the Higgs mixing angle. 
However this bound is not strong as that from K and B meson decays. 

The complex scalar $\chi$ can be stabilized by the remnant $\mathbb{Z}_3$ symmetry after spontaneous symmetry breaking. 
Thus, it can be a dark matter candidate whose mass is given by
\begin{align}
m_{\chi}^2=\frac{1}{2}\left(\mu_\chi^2+\lambda_{H\chi}v^2+\lambda_{\Phi\chi}v_\phi^2\right).
\end{align}

\subsection{The neutrino sector}
The full neutrino mass matrix arises from the Yukawa couplings after spontaneous symmetry breaking: 
\begin{align}
\mathcal{L}_{\nu}^\mathrm{mass}=&
-\overline{\nu_L}m_D\nu_R-\frac{1}{2}\overline{\nu_R^c}M_R\nu_R+\mathrm{h.c.}\nonumber\\
=&~-\left(
\begin{array}{cc}
\overline{\nu_L} & \overline{\nu_R^c} \\
\end{array}
\right)\left(
\begin{array}{cc}
 0 & m_D\\
m_D^T & M_R
\end{array}
\right)\left(
\begin{array}{c}
 \nu_L^c\\
\nu_R
\end{array}
\right)+\mathrm{h.c.}\nonumber\\
\equiv&~-\left(
\begin{array}{cc}
\overline{\nu_L} & \overline{\nu_R^c} \\
\end{array}
\right)
M_{\nu}
\left(
\begin{array}{c}
 \nu_L^c\\
\nu_R
\end{array}
\right)+\mathrm{h.c.},
\end{align}
where $m_D\equiv y_{\nu}v/\sqrt{2}$ and $M_R\equiv y_{\Phi}v_\phi/\sqrt{2}$.
The above $6\times6$ neutrino mass matrix $M_\nu$ can be diagonalized by the unitary matrix $V$ as~\cite{Pilaftsis:1993af}
\begin{align}
 V^{T}M_{\nu}V=\mathrm{diag}(m_1,\cdots,m_6),
\end{align}
and the gauge eigenstates of the neutrinos can be rewritten in terms of the mass eigenstates $\nu_i$ as
\begin{align}
 \left(
\begin{array}{c}
 \nu_L^c\\
 \nu_R
\end{array}
\right)=
V\left(
\begin{array}{c}
 \nu_1^c\\
 \vdots\\
 \nu_6^c
\end{array}
\right),
\hspace{1cm}
 \left(
\begin{array}{c}
 \nu_L\\
 \nu_R^c
\end{array}
\right)=
V^*\left(
\begin{array}{c}
 \nu_1\\
 \vdots\\
 \nu_6
\end{array}
\right),
\end{align}
where $\nu_i~(i=1,2,3)$ corresponds to the active neutrinos with small neutrino masses of $\mathcal{O}(0.1)~\mathrm{eV}$, and $\nu_i~(i=4,5,6)$ are sterile neutrinos with heavy masses. 

The interactions between the Majoron and neutrinos arise from the Yukawa coupling $y_{\Phi}$. 
Using the neutrino mass eigenbasis, the interactions can be written as~\cite{Garcia-Cely:2017oco, Heeck:2019guh} 
\begin{align}
 \mathcal{L}_\nu^\mathrm{int}=&
-\frac{iJ}{2v_\phi}\sum_{i=1}^{6}\sum_{j=1}^{6}
\overline{\nu_i}\left[
C_{ij}\left(m_iP_L-m_jP_R\right)
+C_{ji}\left(m_jP_L-m_iP_R\right)
+\frac{m_i+m_j}{2}\delta_{ij}\gamma_5
\right]\nu_j\nonumber\\
\approx&~\frac{iJ}{2v_\phi}\sum_{i=1}^{6}m_i\overline{\nu_i}\gamma_5\nu_i,
\label{eq:majoron_int}
\end{align}
where the coefficient $C_{ij}$ is defined by $C_{ij}\equiv\sum_{k=1}^{3}V_{ki}V_{kj}^*$, and satisfies the relation $C_{ij}=C_{ji}^*$.
Note that we have taken only the leading term $C_{ij}\approx \delta_{ij}$ in the second line of Eq.~(\ref{eq:majoron_int}).
The Majoron $J$ decays into a pair of neutrinos $J\to \nu_i \nu_i~(i=1,2,3)$ via the interactions, and the decay width $\Gamma_J$ is calculated as
\begin{align}
 \Gamma_J=\sum_{i=1}^{3}\frac{m_J}{16\pi}\frac{m_i^2}{v_\phi^2}.
\label{eq:majoron_decay}
\end{align}
The sum of the active neutrino masses is constrained by the cosmological bound $\sum_{i=1}^{3}m_i<0.11~\mathrm{eV}$~\cite{Planck:2018vyg}. 
In addition, the stronger limit 
\begin{align}
\displaystyle 2.6\times10^{-3}~\mathrm{eV}^2<\sum_{i=1}^{3}m_i^2<5.2\times10^{-3}~\mathrm{eV}^2
\end{align}
is imposed by 
the global fit of the neutrino oscillation data~\cite{deSalas:2020pgw}. 
We take the averaged value $3.9\times10^{-3}~\mathrm{eV^2}$ in the following calculations. 
Successful BBN may be affected if the lifetime of the Majoron is longer than $1~\mathrm{s}$. 
This sets a bound on the parameters as
\begin{align}
\left(\frac{m_J}{100~\mathrm{MeV}}\right)\left(\frac{10~\mathrm{GeV}}{v_\phi}\right)^2 \gtrsim 8.5.
\end{align}

Additional couplings with quarks and charged leptons are induced at the one-loop level~\cite{Pilaftsis:1993af, Garcia-Cely:2017oco}, 
and couplings with gauge bosons are induced at the two-loop level~\cite{Heeck:2019guh}. 
However, these interactions are anticipated to be suppressed sufficiently because of the loop factor and the small Dirac neutrino mass squared $m_Dm_D^\dag$ in the current model. 
These couplings can be important only if the sterile neutrino masses are heavy enough, such as the canonical seesaw scale $M_R\sim10^{14}~\mathrm{GeV}$~\cite{Garcia-Cely:2017oco, Heeck:2019guh}.

\section{Dark matter}
\subsection{Thermal relic abundance}
The complex scalar $\chi$ can be dark matter stabilized by the remnant $\mathbb{Z}_3$ symmetry.
The interactions of the dark matter $\chi$ are induced by the scalar potential.
Considering the plausible mass spectrum $m_J\ll m_\chi\ll m_{h_1,h_2}$, the main dark matter annihilation channels are 
$\chi\overline{\chi}\to JJ$ and the semi-annihilation $\chi\chi\to\overline{\chi}J~(\overline{\chi}\overline{\chi}\to\chi J)$. 
The annihilation cross sections for these channels are calculated as 
\begin{align}
\sigma_{\chi\overline{\chi}}{v}_\mathrm{rel}=&~\frac{1}{16\pi s}\sqrt{1-\frac{4m_J^2}{s}}
\left[
-\frac{\lambda_{H\chi}v}{2v_\phi}\sin\theta\cos\theta\left(
\frac{m_{h_1}^2}{s-m_{h_1}^2}
-
\frac{m_{h_2}^2}{s-m_{h_2}^2}
\right)
\right.\nonumber\\
&~\hspace{3.1cm}
-\frac{\lambda_{H\chi}v}{2v_\phi}\sin\theta\cos\theta\left(m_{h_1}^2-m_{h_2}^2\right)
\left(\frac{\cos^2\theta}{s-m_{h_1}^2}
+
\frac{\sin^2\theta}{s-m_{h_2}^2}
\right)\nonumber\\
&~\hspace{3.1cm}
+\left.\lambda_{\Phi\chi}\left(1+\frac{m_{h_1}^2\sin^2\theta}{s-m_{h_1}^2}+\frac{m_{h_2}^2\cos^2\theta}{s-m_{h_2}^2}\right)\right]^2,
\label{eq:semi0}\\
 \sigma_{\chi\chi}{v}_\mathrm{rel}=&~\sigma_{\overline{\chi}\overline{\chi}}{v}_\mathrm{rel}=
\frac{\lambda^2}{8\pi s}\sqrt{1-\frac{\left(m_\chi+m_J\right)^2}{s}}\sqrt{1-\frac{\left(m_\chi-m_J\right)^2}{s}},
\label{eq:semi}
\end{align}
where $s$ is the Mandelstam variable, and $v_\mathrm{rel}$ is the relative velocity of the dark matter. 
In the above calculation, we included the factor $1/2!$ for identical particles in the final state, 
but not in the initial state in accordance with the standard calculations of quantum field theory~\cite{Peskin:1995ev}. 
The relative sizes of these cross sections were controlled by the quartic couplings $\lambda_{H\chi}$, $\lambda_{\Phi\chi}$, and $\lambda$.  
In this study, we focus on the case of $\lambda_{H\chi},\lambda_{\Phi\chi}\ll\lambda$ because we are interested in the effects of semi-annihilations. 

The number densities of dark matter and anti-dark matter follow the Boltzmann equations:
\begin{align}
 \frac{dn_\chi}{dt}+3Hn_\chi=&
 -\langle\sigma_{\chi\chi}{v}_\mathrm{rel}\rangle\left(n_\chi^2-n_{\overline{\chi}} n_{\chi}^\mathrm{eq}\right)
 +\frac{\langle\sigma_{\overline{\chi}\overline{\chi}}{v}_\mathrm{rel}\rangle}{2}\left(n_{\overline{\chi}}^2-n_{\chi} n_{\chi}^\mathrm{eq}\right),\\
 \frac{dn_{\overline{\chi}}}{dt}+3Hn_{\overline{\chi}}=&
 +\frac{\langle\sigma_{\chi\chi}{v}_\mathrm{rel}\rangle}{2}\left(n_{\chi}^2-n_{\overline{\chi}} n_{\chi}^\mathrm{eq}\right)
 -\langle\sigma_{\overline{\chi}\overline{\chi}}{v}_\mathrm{rel}\rangle\left(n_{\overline{\chi}}^2-n_{\overline{\chi}} n_{\chi}^\mathrm{eq}\right).
\end{align}
Note that the Majoron $J$ is in thermal equilibrium throughout the evolution of (anti-)dark matter number densities in the Boltzmann equations 
because of the moderate Higgs mixing angle. 
Assuming no asymmetry between the dark matter and the anti-dark matter particles ($n_\chi=n_{\overline{\chi}}\equiv n/2$),\footnote{This assumption is reasonable because there is no CP violation in the scalar potential.} 
the total number density of the dark matter and anti-dark matter particles defined by $n$ follows the combined Boltzmann equation:
\begin{align}
 \frac{dn}{dt}+3Hn=&-\frac{\langle\sigma_{\chi\chi}{v}_\mathrm{rel}\rangle}{4}\left(n^2-n n^\mathrm{eq}\right).
\end{align}
Note that the factor $1/4$ appears on the right-hand side compared with the case of the canonical WIMPs because we consider the semi-annihilation 
of the complex scalar dark matter here~\cite{Gondolo:1990dk}.
Although the Boltzmann equation is numerically solved to reproduce the observed relic abundance $\Omega h^2\approx0.12$ in the later part of the paper~\cite{Planck:2018vyg}, 
the semi-analytic solution can also easily be obtained~\cite{DEramo:2010keq}. 
The semi-annihilation cross section in Eq.~(\ref{eq:semi}) can be simplified as $\sigma_{\chi\chi}v_\mathrm{rel}\approx3\lambda^2/(128\pi m_\chi^2)$ in $m_J\ll m_\chi$ and the non-relativistic limit. 
Then, because the typical magnitude of the required cross section is $\sigma_{\chi\chi}v_\mathrm{rel}\sim 2\times10^{-25}~\mathrm{cm^3/s}$ for the semi-annihilating complex scalar dark matter, 
one can obtain the required magnitude of the quartic coupling $\lambda$ as
\begin{align}
\lambda\approx 1.6\times10^{-3}\left(\frac{m_\chi}{\mathrm{GeV}}\right),
\end{align}
to reproduce the observed relic abundance.

\subsection{Halo core formation via semi-annihilation}
It has been proposed that semi-annihilating dark matter can form cores of dark matter halos because the dark matter mass is converted into kinetic energy via semi-annihilation, 
and the dark matter particles in the inner region of the halos are thermalized with pressure~\cite{Chu:2018nki}. 
This effect alleviates small-scale problems such as core-cusp, too-big-to-fail, and missing satellite problems~\cite{Tulin:2017ara}. 
Core formation via semi-annihilating dark matter occurs more frequently for dwarf-sized halos than for more massive halos in the same time scale. 
This is a different feature from core formation via strongly self-interacting dark matter~\cite{Tulin:2017ara}. 
Another feature of the core formation via semi-annihilating dark matter is dark matter velocity dependence, as we show in the following.

The order of the semi-annihilation cross section required for core formation was estimated as~\cite{Chu:2018nki}
\begin{align}
\langle\sigma_{\chi\chi}{v}_\mathrm{rel}\rangle=
\frac{10^{-29}~\mathrm{cm^3/s}}{\xi}
\left(\frac{m_\chi}{100~\mathrm{MeV}}\right)
\left(\frac{\sigma_0}{5~\mathrm{km/s}}\right)^2
\left(\frac{0.5~M_\odot~\mathrm{pc}^{-3}}{\rho_c}\right)
\left(\frac{10^{10}~\mathrm{yr}}{t_\mathrm{age}}\right),
\end{align}
where $\xi$ is the energy absorption efficiency given by the ratio between the dark matter halo radius and the mean free path of the dark matter
\begin{align}
\xi=10^{-3}\left(\frac{r_s}{5~\mathrm{kpc}}\right)
\left(\frac{\rho_c}{M_\odot~\mathrm{pc}^{-3}}\right)
\left(\frac{\sigma_\mathrm{self}/m_\chi}{10^{-3}~\mathrm{cm^2/g}}\right),
\label{eq:xi}
\end{align}
$\sigma_0$ is the dark matter velocity dispersion due to thermalization, $\rho_c$ is the dark matter core density of the halo, 
$r_s$ is the halo radius, $M_\odot$ is the solar mass, $t_\mathrm{age}$ is the age of the universe, and $\sigma_\mathrm{self}$ is the self-interacting cross section of dark matter. 
Note that the order of the self-interacting cross section required for core formation is typically $\sigma_\mathrm{self}/m_\chi\sim10^{-3}~\mathrm{cm^2/g}$, 
which is much smaller than the case of strongly self-interacting dark matter requiring $\mathcal{O}(1)~\mathrm{cm^2/g}$~\cite{Tulin:2017ara}.

In the current model, the self-interacting cross sections can be calculated as~\cite{Choi:2015bya}
\begin{align}
 \sigma_{\chi\chi\to\chi\chi}=&~\sigma_{\overline{\chi}\overline{\chi}\to\overline{\chi}\overline{\chi}}=
\frac{1}{32\pi m_\chi^2}\left(\lambda_\chi+\frac{\lambda^2 v_\phi^2}{6m_\chi^2}\right)^2,\\
\sigma_{\chi\overline{\chi}\to\chi\overline{\chi}}=&~
\frac{1}{16\pi m_\chi^2}\left(\lambda_\chi-\frac{\lambda^2v_\phi^2}{2m_\chi^2}\right)^2,
\end{align}
in non-relativistic limit. 
Although additional contributions are induced by the couplings $\lambda_{H\chi}$ and $\lambda_{\Phi\chi}$, these contributions are subdominant 
owing to the current setup $\lambda_{H\chi},\lambda_{\Phi\chi}\ll\lambda$. 
Because there is no asymmetry between the dark matter and the anti-dark matter particles, 
the effective self-interacting cross section can be defined by
\begin{align}
 \sigma_\mathrm{self}^\mathrm{eff}\equiv&~
\frac{1}{4}\sigma_{\chi\chi\to\chi\chi}+\frac{1}{4}\sigma_{\chi\overline{\chi}\to\chi\overline{\chi}}
+\frac{1}{4}\sigma_{\overline{\chi}\overline{\chi}\to\overline{\chi}\overline{\chi}}\nonumber\\
=&~\frac{1}{64\pi m_\chi^2}\left[\left(\lambda_\chi+\frac{\lambda^2v_\phi^2}{6m_\chi^2}\right)^2+\left(\lambda_\chi-\frac{\lambda^2v_\phi^2}{2m_\chi^2}\right)^2\right],
\end{align}
and quantity $\sigma_\mathrm{self}/m_\chi$ in Eq.~(\ref{eq:xi}) should be replaced by the effective cross section $\sigma_\mathrm{self}^\mathrm{eff}/m_\chi$.

\subsection{Perturbative unitarity bound}
\begin{figure}[t]
\begin{center}
\includegraphics[width=8cm]{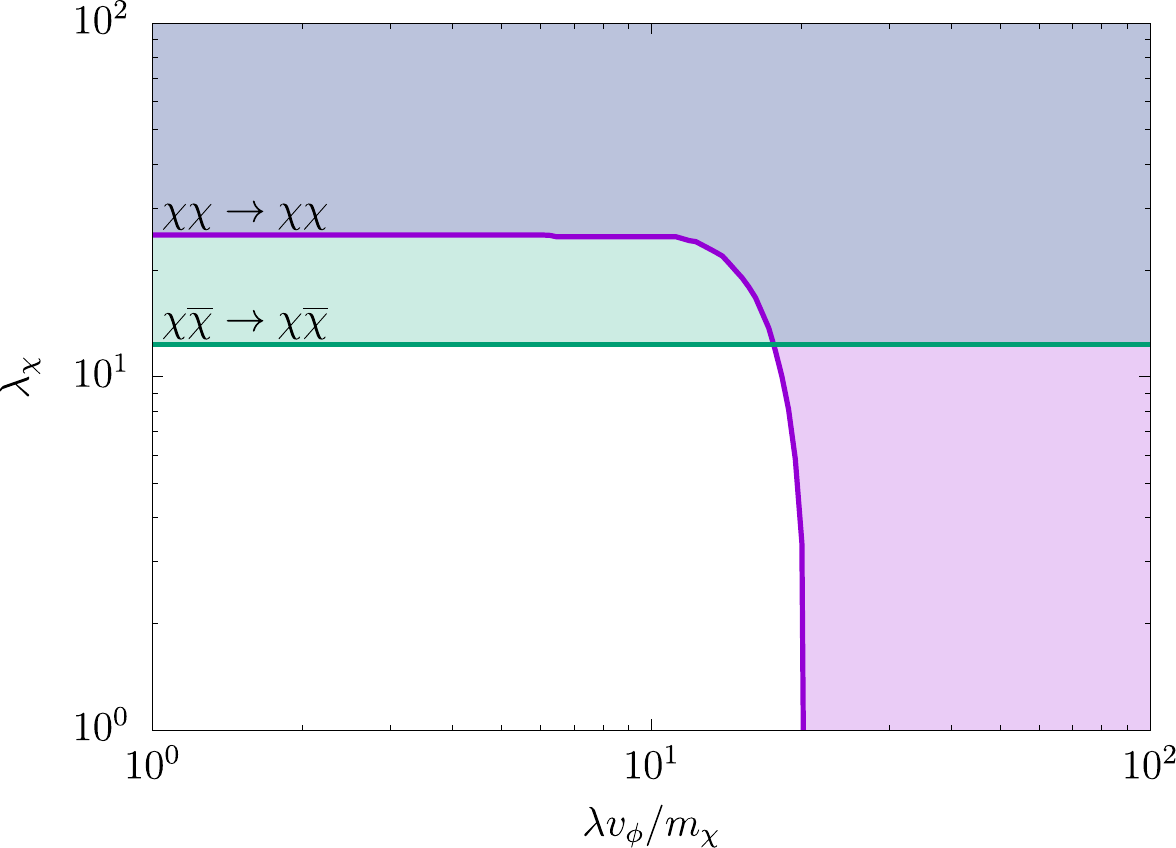}
\caption{Unitarity bounds obtained from the scattering processes $\chi\chi\to\chi\chi$ and $\chi\overline{\chi}\to\chi\overline{\chi}$. Colored regions are excluded.}
\label{fig:uni}
\end{center}
\end{figure}

The perturbativity bounds for the quartic couplings $\lambda$ and $\lambda_\chi$ are simply given by $\lambda,\lambda_\chi<4\pi$. 
The unitarity bounds can also be imposed by the calculation of the scattering amplitudes for $\chi\chi\to\chi\chi$ and $\chi\overline{\chi}\to\chi\overline{\chi}$, which are given by~\cite{DiLuzio:2016sur, Goodsell:2018tti}
\begin{align}
\sqrt{1-\frac{4m_\chi^2}{s}}\left[
2\lambda_\chi+\frac{\lambda^2v_\phi^2}{s-m_\chi^2}
\right]
&\leq 16\pi,
\label{eq:uni1}\\
\sqrt{1-\frac{4m_\chi^2}{s}}\left[
2\lambda_\chi
-\frac{\lambda^2v_\phi^2}{s-4m_\chi^2}\log\left(\frac{s-3m_\chi^2}{m_\chi^2}\right)
\right]&\leq 8\pi,
\label{eq:uni2}
\end{align}
respectively. 
In the high-energy limit $s\to\infty$, the bounds are simplified to the usual ones $\lambda_\chi\leq 8\pi$ and $\lambda_\chi\leq 4\pi$ 
and, thus, no unitarity bound for the coupling $\lambda$. 
However, the above unitarity bounds in Eqs.~(\ref{eq:uni1}) and (\ref{eq:uni2}) should be satisfied in all ranges of the Mandelstam variable $s$. 
Thus, we vary $s$ in the range of $4m_\chi^2\lesssim s<\infty$ to obtain a conservative bound for the coupling $\lambda$.
The results are presented in Fig.~\ref{fig:uni}, and we find that the bound $\lambda v_\phi/m_\chi\lesssim17.5$ is numerically obtained from the figure.

\subsection{Direct detection}
The dark matter $\chi$ can scatter a nucleon off via the couplings $\lambda_{H\chi}$ and $\lambda_{\Phi\chi}$ in the scalar potential. 
The constraint of dark matter direct detection is not so severe because the dark matter mass scale is sub-GeV in our setup as we will see later.
However for completeness, we evaluate the upper bound on these couplings below. 
The elastic scattering cross section between dark matter and nucleon at zero momentum transfer is calculated as
\begin{align}
\sigma_{\chi N}=\frac{m_N^4v_\phi^4f_N^2\left(\lambda_{H\chi}\lambda_\Phi-\lambda_{\Phi\chi}\lambda_{H\Phi}\right)^2}{4\pi(m_\chi+m_N)^2m_{h_1}^4m_{h_2}^4},
\end{align}
where $f_N=\sum_{q}f_N^q$ is the scalar quark form factors, which can be evaluated as $f_p=$0.284 for a proton and $f_n=0.287$ for a nucleon~\cite{Belanger:2018ccd}. 
Taking the upper bound on the elastic cross section $\sigma_\mathrm{el}^\mathrm{exp}\sim10^{-40}~\mathrm{cm}^2$ at $m_{\chi}=1~\mathrm{GeV}$~\cite{ParticleDataGroup:2020ssz}, 
we can derive the bound 
\begin{align}
\frac{v_\phi^4}{m_{h_2}^4}\left(\lambda_{H\chi}\lambda_\Phi-\lambda_{\Phi\chi}\lambda_{H\Phi}\right)^2\lesssim0.011.
\end{align}
one can see that this bound is not strong because of the relation $v_\phi\lesssim m_{h_2}$. 
The future direct detection experiment SuperCDMS is sensitive to sub-GeV scale dark matter and is anticipated to update the bound up to $10^{-43}~\mathrm{cm}^2$~\cite{SuperCDMS:2016wui}.

\subsection{Box-shaped neutrino spectra}

\begin{figure}[t]
\begin{center}
 \includegraphics[width=4cm]{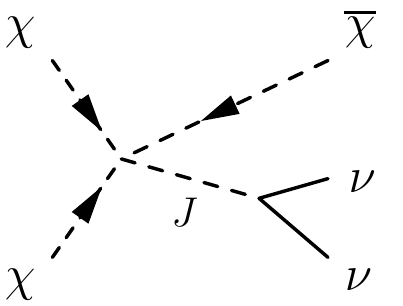}
\caption{Diagram of the neutrino production from the semi-annihilation.}
\label{fig:semi-ann}
\end{center} 
\end{figure}

\begin{figure}[t]
\begin{center}
\includegraphics[width=8cm]{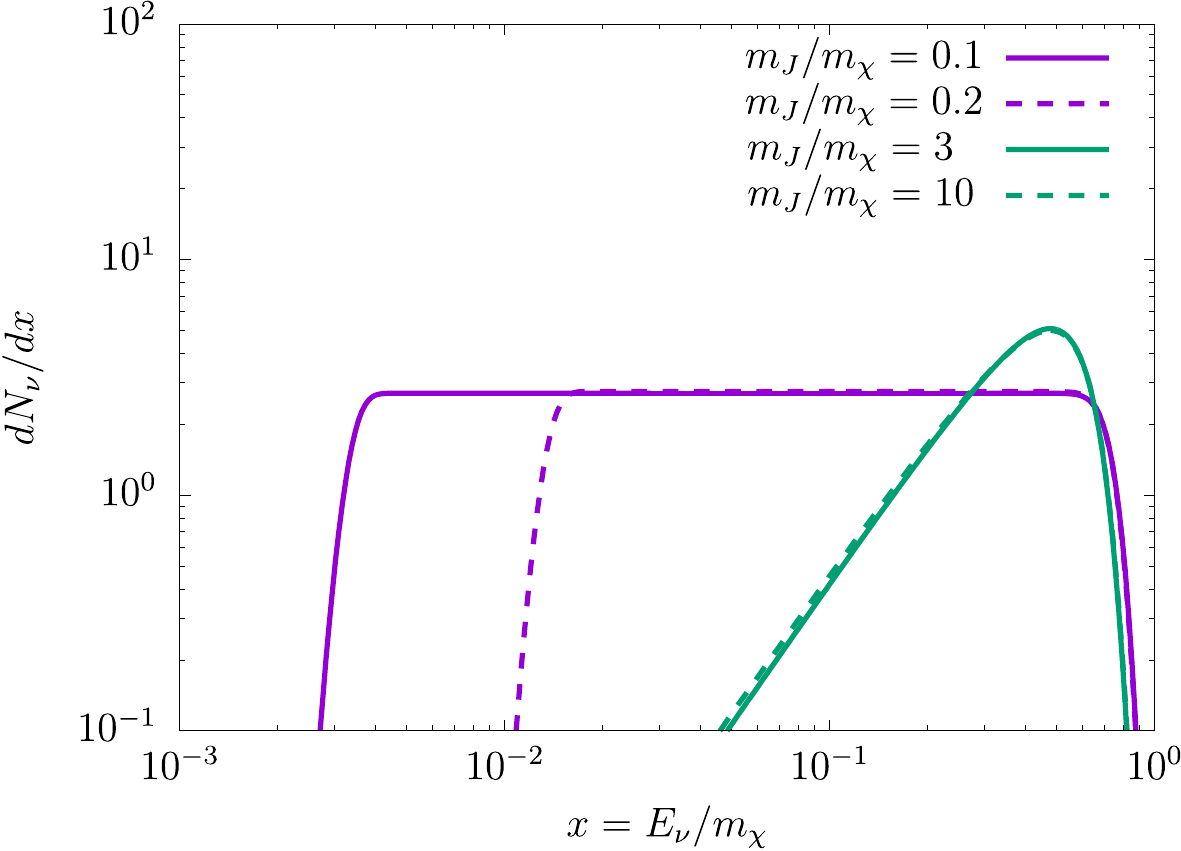}
\caption{Energy spectra of the neutrino $\nu$ produced by the semi-annihilation processes where 
the energy resolution of $\sigma/x=10\%$ is taken into account.
The purple lines represent the spectra for the on-shell Majoron $J$ ($m_J\ll m_\chi$), whereas the green lines represent the spectra for the off-shell Majoron $J$ ($m_\chi<m_J$). 
See Section~\ref{sec:off-shell} for the off-shell case.}
\label{fig:dndx}
\end{center} 
\end{figure}

The semi-annihilation processes $\chi\chi\to\overline{\chi}J$ ($\overline{\chi}\overline{\chi}\to\chi J$) produce the on-shell Majoron $J$.
Then, the neutrinos are produced by the subsequent decay of the Majoron $J\to\nu_i\nu_i~(i=1,2,3)$, as shown in Fig.~\ref{fig:semi-ann}. 
These neutrinos can be a signal of semi-annihilating dark matter in the model.
The energy of the (anti-)dark matter particle in the final state and the on-shell Majoron are kinematically fixed by the energy--momentum conservation as
\begin{align}
E_{\overline{\chi}}=\frac{5m_\chi^2-m_J^2}{4m_\chi},\quad
E_{J}=\frac{3m_\chi^2+m_J^2}{4m_\chi},
\end{align}
because the dark matter particles in the initial state are non-relativistic. 
The differential decay width of the neutrinos can be computed as~\cite{Ibarra:2012dw}
\begin{align}
 \frac{d\Gamma_J}{dE_{\nu}}=
\frac{\Gamma_J}{\sqrt{E_J^2-m_J^2}}\theta\big(E_{\nu}-E_{\nu}^{\mathrm{min}}\big)\theta\big(E_{\nu}^{\mathrm{max}}-E_{\nu}\big),
\end{align}
where the minimum and maximum neutrino energy $E_\nu^\mathrm{min}$ and $E_\nu^\mathrm{max}$ are given by
\begin{align}
 E_{\nu}^\mathrm{min}=\frac{E_J}{2}\left(1-\sqrt{1-\frac{m_J^2}{E_J^2}}\right),\quad
 E_{\nu}^\mathrm{max}=\frac{E_J}{2}\left(1+\sqrt{1-\frac{m_J^2}{E_J^2}}\right),
\label{eq:emax}
\end{align}
and the total decay width of Majoron $\Gamma_J$ is given by Eq.~(\ref{eq:majoron_decay}). 
Therefore, the produced neutrino energy spectrum becomes box-shaped, as shown by the purple lines in Fig.~\ref{fig:dndx}, where 
the neutrino energy spectrum $dN_{\nu}/dx$ is defined by 
\begin{align}
 \frac{dN_{\nu}}{dx}\equiv\frac{2}{\Gamma_J}\frac{d\Gamma_J}{dx},
\end{align}
with the dimensionless parameter $x\equiv E_{\nu}/m_\chi$. 
The factor $2$ comes from the fact that two neutrinos are produced for each Majoron decay.
In Fig.~\ref{fig:dndx}, the energy resolution of the experimental detectors $\sigma$ was considered by~\cite{Ibarra:2012dw}
\begin{align}
 \frac{dN_{\nu}}{dx}=\int \frac{dN_{\nu}}{dx^\prime}\frac{1}{\sqrt{2\pi}\sigma}e^{-(x^\prime-x)^2/2\sigma^2}dx^\prime.
\end{align}

In the case of semi-annihilating dark matter, the neutrino flux produced by dark matter semi-annihilations in the Milky Way halo is given by~\cite{Olivares-DelCampo:2017feq}
\begin{align}
 \frac{d\Phi_\nu}{dx}=\frac{\langle\sigma_{\chi\chi}v_\mathrm{rel}\rangle}{4}\mathcal{J}_\mathrm{avg}
\frac{R_0\rho_0^2}{m_\chi^2}\frac{dN_{\nu}}{dx},
\end{align}
where $\mathcal{J}_\mathrm{avg}$ is the $J$-factor averaged over the Milky Way halo, $R_0=8.5~\mathrm{kpc}$ is the distance between the galactic center and the solar system, 
and $\rho_0=0.3~\mathrm{GeV/cm^3}$ is the local dark matter density in the solar system.
Although the $J$-factor generally depends on the chosen dark matter halo profiles, it is not sensitive because the $J$-factor is averaged over a large region of the solid angle~\cite{Yuksel:2007ac}. 
Using data from Super-Kamiokande (SK) and the future sensitivity of HK, 
the current bound and future prospects on the standard dark matter annihilation cross section for the channel $\chi\chi\to\nu\nu$ in the mass range 
$10~\mathrm{MeV}\lesssim m_{\chi}\lesssim200~\mathrm{MeV}$ 
have been studied with the averaged $J$-factor $\mathcal{J}_\mathrm{avg}=5$ in~\cite{Olivares-DelCampo:2017feq, Olivares-DelCampo:2018pdl}. 
These bound and future prospects for the monochromatic neutrino lines can be translated into the case of the box-shaped spectrum in our case by comparing 
the predicted box-shaped neutrino flux at $E_{\nu}=E_\nu^\mathrm{max}$, given by Eq.~(\ref{eq:emax}) to the monochromatic flux at $E_\nu=m_\chi$. 
This is plausible because the predicted neutrino flux at $E_{\nu}=E_\nu^\mathrm{max}$ is expected to provide the strongest limit on the cross section in our model.

\subsection{Off-shell Majoron case}
\label{sec:off-shell}
Here, we consider the case of $m_\chi<m_J\ll m_{h_2}$ for completeness, although it may be less motivated from the viewpoint of the light pseudo-Nambu--Goldstone boson. 
In this case, the semi-annihilation $\chi\chi\to\overline{\chi}J$ and its CP conjugate processes are kinematically forbidden. 
However, three-body semi-annihilation $\chi\chi\to\overline{\chi}\nu\nu$ is possible via the off-shell Majoron.
The differential cross section is calculated as
\begin{align}
\frac{d\sigma_{\chi\chi}{v}_\mathrm{rel}}{dx}=
\frac{\lambda^2\Gamma_J}{16\pi^2 m_\chi^3m_J}
\left[\frac{z}{1-z}+\log(1-z)\right],
\label{eq:sv_off}
\end{align}
where the dimensionless parameter $z$ is defined as $z\equiv wx\left(3-4x\right)/(1-x)$ with $w=m_\chi^2/m_J^2<1$.
The total cross section can be obtained by integrating Eq.~(\ref{eq:sv_off}) in the range $0<x<3/4$,
\begin{align}
\sigma_{\chi\chi}{v}_\mathrm{rel}=&~\frac{\lambda^2m_J\Gamma_J}{64\pi^2 m_\chi^4}
\Bigg[
\frac{2-15w+9w^2}{2\sqrt{(1-w)(1-9w)}}
\log\left(\frac{5-9w-3\sqrt{(1-w)(1-9w)}}{5-9w+3\sqrt{(1-w)(1-9w)}}\right)\nonumber\\
&~\hspace{1.7cm}
-6w+(2-5w)\log{2}
\Bigg].
\label{eq:sv_off2}
\end{align}
Based on the above calculation, the neutrino spectrum defined by
\begin{align}
 \frac{dN_\nu}{dx}\equiv\frac{2}{\sigma_{\chi\chi}v_\mathrm{rel}}\frac{d\sigma_{\chi\chi}v_\mathrm{rel}}{dx},
\end{align}
is shown as the green lines in Fig.~\ref{fig:dndx}.
The spectra can be significantly larger than the box-shaped spectra at high energies. 
However, the bound and future prospects for the cross section are much weaker than the on-shell Majoron case studied above because the cross section in Eq.~(\ref{eq:sv_off2}) is suppressed by 
the small decay width of the Majoron $\Gamma_J$ given by Eq.~(\ref{eq:majoron_decay}). 
Furthermore, this suppressed cross section was too small to reproduce the observed relic abundance.\footnote{The other ways to detect the neutrinos for the off-shell case at the DUNE experiment and colliders have been studied though these cannot be simply applied to our case~\cite{Kelly:2019wow, Kelly:2021mcd}.} 

Therefore, the relic abundance of dark matter is determined by three-to-two self-annihilation processes 
such as $\chi\chi\chi\to\chi\overline{\chi}$ and $\chi\overline{\chi}\chi\to\overline{\chi}\overline{\chi}$, instead of semi-annihilations.\footnote{Another three-to-two annihilation channel $\chi\overline{\chi}\chi\to \chi J$ opens if $m_\chi \lesssim m_J\lesssim2m_\chi$.} 
Assuming no CP asymmetry in the dark sector, as in the previous case, the Boltzmann equation for the total dark matter number density $n$ is given by~\cite{Choi:2015bya}
\begin{align}
 \frac{dn}{dt}+3Hn=-\frac{1}{4}\langle\overline{\sigma}_{3\to2}v_\mathrm{rel}^2\rangle\left(n^3-n^2n_\mathrm{eq}\right),
\end{align}
where $\langle\overline{\sigma}_{3\to2}v_\mathrm{rel}^2\rangle$ is the effective three-to-two cross section defined by
\begin{align}
\langle\overline{\sigma}_{3\to2}v_\mathrm{rel}^2\rangle\equiv&~
\frac{1}{3!}\langle\sigma_{\chi\chi\chi}v_\mathrm{rel}^2\rangle+\frac{1}{2!}\langle\sigma_{\chi\overline{\chi}\chi}v_\mathrm{rel}^2\rangle\nonumber\\
=&~\frac{\sqrt{5}\lambda^2v_\phi^2}{1152\pi m_\chi^7}\left[\frac{1}{3!}\left(\frac{\lambda^2v_\phi^2}{2m_\chi^2}+\lambda_\chi\right)^2
+\frac{1}{2!}\frac{1}{48}\left(\frac{13\lambda^2v_\phi^2}{2m_\chi^2}-37\lambda_\chi\right)^2\right].
\label{eq:3to2}
\end{align}
In Eq.~(\ref{eq:3to2}), the thermally averaged cross sections $\langle\sigma_{\chi\chi\chi}v_\mathrm{rel}^2\rangle$ and $\langle\sigma_{\chi\overline{\chi}\chi}v_\mathrm{rel}^2\rangle$ 
include the symmetry factors for identical particles in the final state, but not in the initial state as same as Eqs.~(\ref{eq:semi0}) and (\ref{eq:semi}), and 
only the $s$-wave contribution is taken into account because the dark matter particles in the initial state are non-relativistic.

The analytic solution for the observed relic abundance is approximately given by 
$m_\chi\approx0.035\alpha_\mathrm{eff}$, where the effective coupling $\alpha_\mathrm{eff}$ is defined by $\langle\overline{\sigma}_{3\to2}v_\mathrm{rel}^2\rangle\equiv\alpha_\mathrm{eff}^3/m_\chi^5$~\cite{Hochberg:2014dra, Hochberg:2014kqa, Choi:2015bya}. 
Thus, for the simple case with $\lambda_\chi=0$, because the effective coupling is evaluated as $\alpha_\mathrm{eff}\approx 0.034\lambda^2v_\phi^2/m_\chi^2$, we obtain the solution $\lambda\approx29\left(m_\chi/\mathrm{GeV}\right)^{1/2}\left(m_\chi/v_\phi\right)$.

\section{Numerical results}

We perform numerical computation and find the parameter space that can simultaneously reproduce the observed dark matter relic abundance, 
form dark matter halo cores and exhibit a neutrino signal from the Majoron decay. 
We have five parameters relevant to the numerical computation: $m_\chi$, $m_J$, $v_\phi$, $\lambda$, and $\lambda_\chi$. 
The Majoron mass should be much smaller than the dark matter mass so that sufficient kinetic energy is provided to the (anti-)dark matter particles by the semi-annihilation processes, and the neutrino energy spectrum becomes box-shaped. 
Here we take $m_J/m_\chi=0.1$ as a sample point. 
Note that the precise value of the Majoron mass does not practically affect the relic abundance and neutrino signals 
as long as $m_J/m_\chi\ll1$, whereas the decay width of the Majoron can be affected by Eq.~(\ref{eq:majoron_decay}). 

\begin{figure}[t]
\begin{center}
\includegraphics[width=10cm]{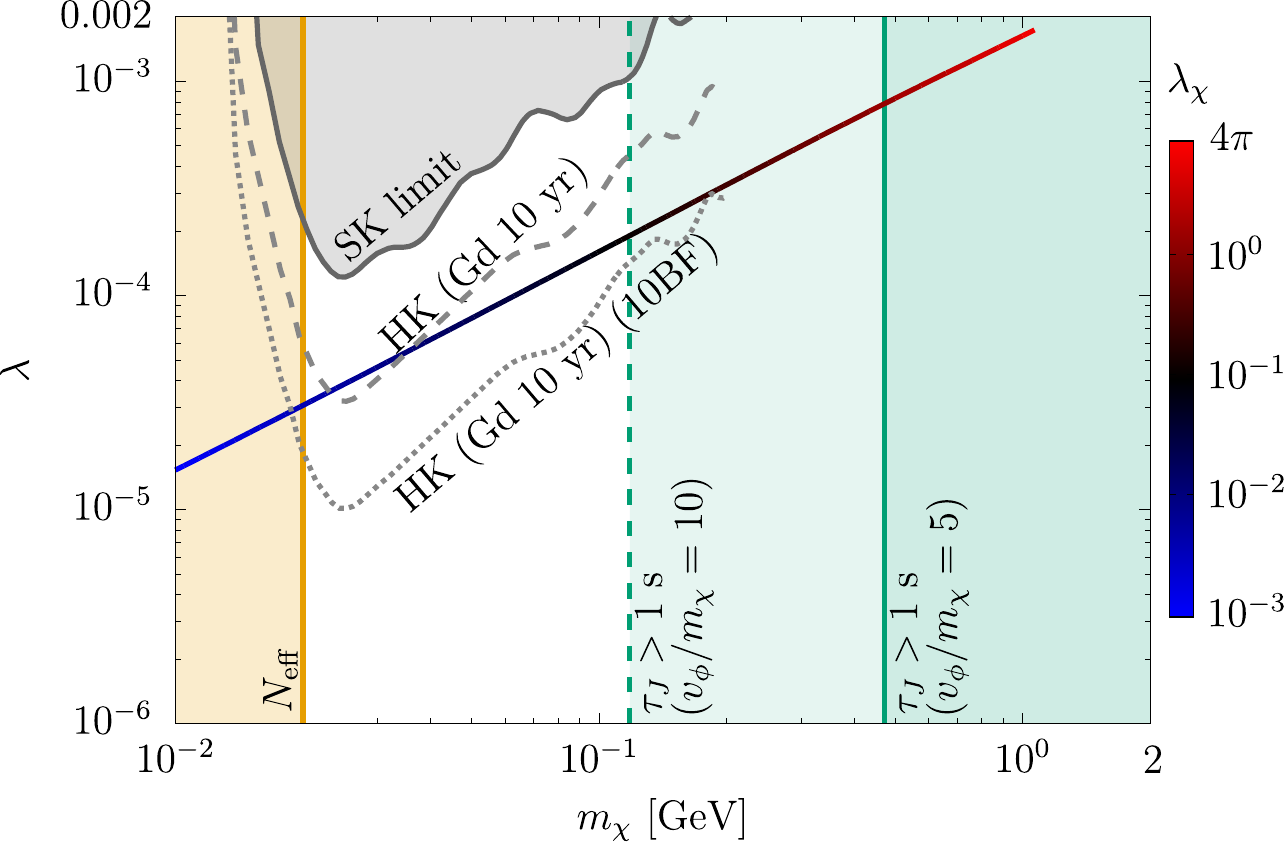}
\caption{Parameter space in the ($m_\chi,\lambda$) plane with $m_J/m_\chi=0.1$. 
The gradient line from blue, black, and red represents the parameter space that can reproduce the observed dark matter relic abundance with the semi-annihilation $\chi\chi\to\overline{\chi}J$ 
where the value of the quartic coupling $\lambda_\chi$ required for dark matter core formation changes from $10^{-3}$ (blue) to $4\pi$ (red). 
The orange region on the left-hand side is ruled out by $N_\mathrm{eff}$ (freeze-out temperature $\lesssim1~\mathrm{MeV}$). 
The green regions are excluded by the lifetime of the Majoron ($\tau_J>1~\mathrm{s}$) for a fixed $v_\phi/m_\chi=5,10$. 
The gray region is also excluded by the SK data~\cite{Olivares-DelCampo:2017feq}. 
The dashed and dotted gray lines represent the HK sensitivity using Gd 10 years and 
without/with a boost factor of 10 (10BF) at the galaxies compared with the value required for the relic abundance~\cite{Olivares-DelCampo:2018pdl}. 
}
\label{fig:mx_la}
\end{center} 
\end{figure}

The numerical result is shown in Fig.~\ref{fig:mx_la}. 
In the orange region on the left-hand side ($m_\chi\lesssim20~\mathrm{MeV}$), the dark matter $\chi$ is still in a thermal bath at the temperature $\sim1~\mathrm{MeV}$ and 
affects the effective number of neutrino species $N_\mathrm{eff}$~\cite{Boehm:2012gr, Boehm:2013jpa}. 
In the green region on the right-hand side, the successful BBN is spoiled by the additional energy injection to the SM sector through $J\to\nu\nu$ because the Majoron is 
too long-lived ($\tau_J>1~\mathrm{s}$), as shown in Eq.~(\ref{eq:majoron_decay}). 
The relevant parameter space depends on the VEV $v_\phi$, and we have taken $v_\phi/m_\chi=5,10$ in Fig.~\ref{fig:mx_la}. 
The straight line whose color gradually changes from blue to red corresponds to the parameter space that can reproduce the observed relic abundance by semi-annihilation. 
The color represents the value of the quartic coupling $\lambda_\chi$ required for the core formation of dark matter halos. 
The perturbative unitarity bound for the quartic coupling is imposed $\lambda_\chi<4\pi$. 
From Fig.~\ref{fig:mx_la}, we can see that the dark matter mass is $m_\chi\lesssim\mathcal{O}(1)~\mathrm{GeV}$ 
to reproduce the observed relic abundance and realize the core formation of dark matter halos via semi-annihilation simultaneously. 

The gray region was excluded by the neutrino observation from the galactic center by the SK~\cite{Olivares-DelCampo:2017feq}. 
The dashed gray line represents the future prospects of the HK with gadolinium 10 years observation~\cite{Olivares-DelCampo:2018pdl}. 
One can find that the sensitivity of the HK can reach the parameter space favored for dark matter relic abundance and core formation 
if the dark matter mass is in the range of $20~\mathrm{MeV}\lesssim m_\chi\lesssim 35~\mathrm{MeV}$. 
The lower dotted gray line is the future prospect with a boost factor of $10$ for the semi-annihilation cross section. 
Such enhancement of the annihilation cross section can be realized by mechanisms such as the Breit--Wigner effect~\cite{Griest:1990kh, Ibe:2008ye} and 
the Sommerfeld effect~\cite{Hisano:2002fk, Hisano:2003ec, Hisano:2004ds, Hisano:2005ec, Hisano:2006nn}. 
For example, the simplest extension of the model can be achieved by introducing a singlet scalar $S$ with a global $U(1)_{B-L}$ charge $-4/3$. 
Then, the semi-annihilation channel $\chi\chi\to S^* \to \overline{\chi}J$ is enhanced at the resonance at $2m_\chi\approx m_S$. 
If this kind of enhancement occurs, the dark matter mass up to $m_\chi\sim200~\mathrm{MeV}$ can be tested by HK with Gd, as shown in Fig.~\ref{fig:mx_la}.

\begin{figure}[t]
\begin{center}
\includegraphics[width=7.9cm]{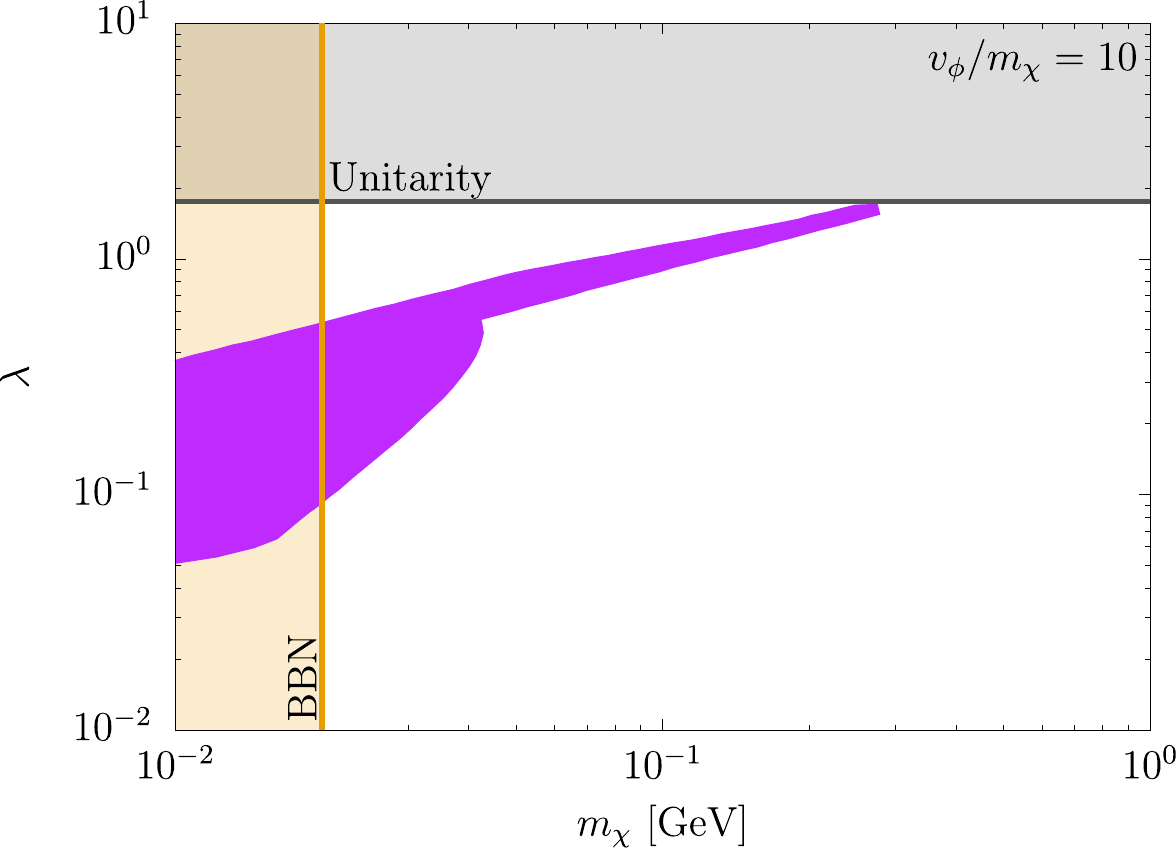}
\includegraphics[width=7.9cm]{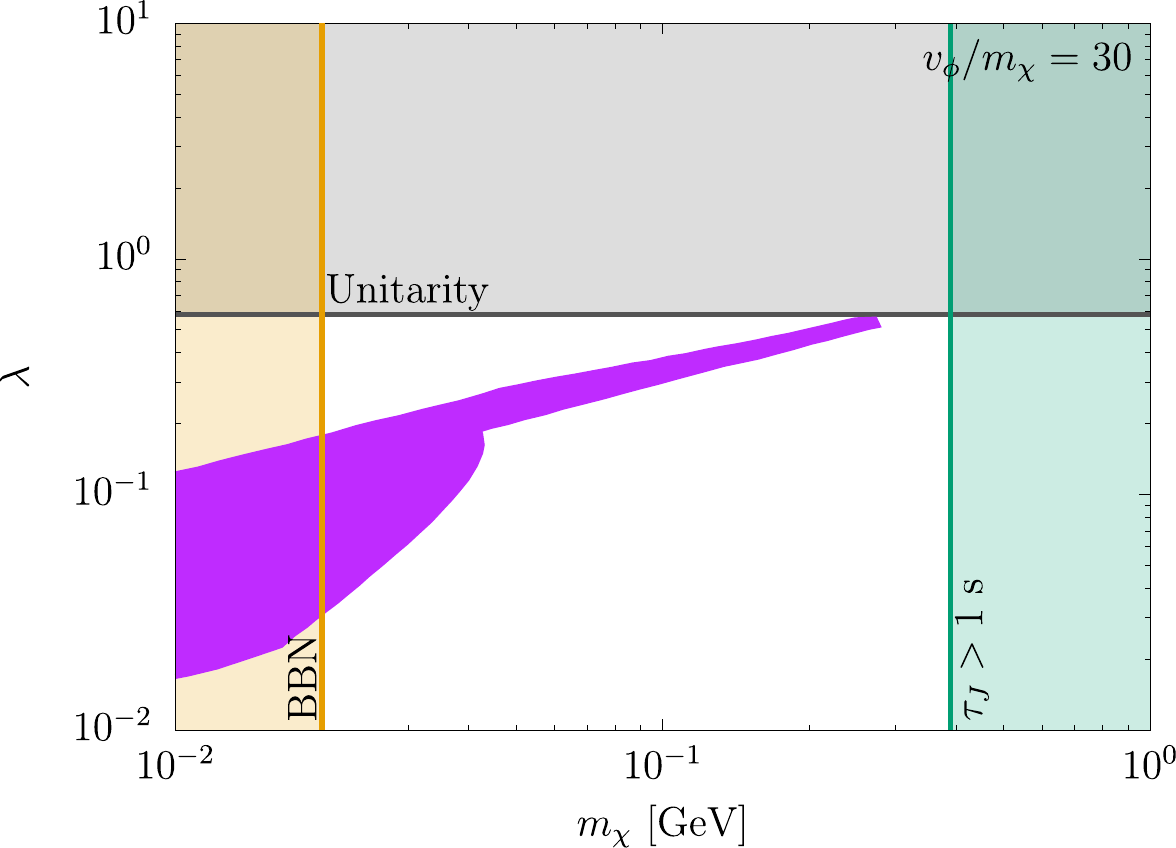}
\caption{Parameter space in the ($m_\chi,\lambda$) plane where the Majoron mass is fixed to be $m_J/m_\chi=3$.
The purple region can reproduce the observed relic abundance and form dark matter cores via strongly self-interacting dark matter. 
($0.1~\mathrm{cm^2/g}\leq\sigma_\mathrm{self}/m_\chi\leq1~\mathrm{cm^2/g}$). 
The gray region is excluded by the unitarity bound for the quartic coupling $\lambda$, and the other colored regions are the same as in Fig.~\ref{fig:mx_la}.}
\label{fig:simp}
\end{center} 
\end{figure}

The numerical results for the off-shell Majoron case are shown in Fig.~\ref{fig:simp}. 
The purple region can reproduce the observed relic abundance and form dark matter halo cores simultaneously. 
Here, we have taken the self-interacting cross section required for the core formation as $0.1~\mathrm{cm^2/g}\leq\sigma_\mathrm{self}/m_\chi\leq1~\mathrm{cm^2/g}$~\cite{Tulin:2017ara}. 
The upper gray region is excluded by the unitarity bound for quartic coupling $\lambda$. 
The other colored regions were also excluded for the same reasons as in Fig.~\ref{fig:mx_la}. 
For $m_\chi\gtrsim40~\mathrm{MeV}$, the purple region approximately coincides with the analytical solution for the relic abundance $\lambda\approx29\left(m_\chi/\mathrm{GeV}\right)^{1/2}\left(m_\chi/v_\phi\right)$. 
In this mass region, the quartic coupling $\lambda$ is dominant over the other coupling $\lambda_\chi$ to determine the relic abundance and form the dark matter core formation. 
For the smaller mass region, because these couplings are comparable or $\lambda_\chi$ is dominant, a wider region of $\lambda$ can be chosen at a fixed dark matter mass. 
Compared with the semi-annihilation case ($m_J<m_\chi$), one finds that the quartic coupling $\lambda$ is closer to the unitarity bound because a stronger self-interacting cross section is required for dark matter halo core formation.

\section{Summary and discussion}
Although the canonical WIMPs are strongly constrained by recent experimental and observational results, the thermal production mechanism of dark matter is still attractive. 
Semi-annihilating dark matter is an alternative to thermal dark matter, and semi-annihilations indicate some interesting phenomenological aspects 
which do not emerge for the standard annihilations of dark matter, such as core formation of dark matter halos, and boosted dark matter in the final states. 
In this paper, we have proposed a model of semi-annihilating dark matter based on global $U(1)_{B-L}$ symmetry, where two complex scalars and 
three right-handed neutrinos are introduced as new fields to the SM. 
The complex scalar is stabilized by the remnant $\mathbb{Z}_3$ symmetry after spontaneous symmetry breaking. 
In addition, the small neutrino masses are also generated via the seesaw mechanism. 
Because the pseudo-Nambu--Goldstone boson associated with the global $U(1)_{B-L}$ symmetry, the so-called Majoron, is anticipated to be light enough, 
the dark matter semi-annihilation into the Majoron is naturally regarded as the dominant channel compared with the other annihilation channels.

We have explored a parameter space that can reproduce the observed relic abundance via the canonical freeze-out mechanism, form dark matter halo cores, and 
be consistent with the cosmological observations and the perturbative unitarity bounds. 
We have found that the dark matter mass should be in the range of $20~\mathrm{MeV}\lesssim m_\chi\lesssim 1~\mathrm{GeV}$, and the quartic coupling $\lambda$ that induces semi-annihilation should be in the range of $3\times10^{-5}\lesssim \lambda\lesssim 2\times10^{-3}$. 

In addition, the box-shaped spectrum of neutrinos can be generated by semi-annihilation in the galaxies as a characteristic signature of the model. 
The signal can be detected at the future large-volume neutrino detector, HK, if the dark matter mass is rather light, $25~\mathrm{MeV}\lesssim m_\chi\lesssim 35~\mathrm{MeV}$. 
Furthermore, a larger dark matter mass region can be tested if the semi-annihilation cross section is enhanced by a mechanism such as the Breit--Wigner and Sommerfeld effects. 

For comparison, we have also investigated the case in which the Majoron mass is heavier than the dark matter mass. 
In this case, the dark matter relic abundance is determined by the three-to-two annihilation processes rather than the semi-annihilation processes. 
As a result, stronger self-interactions are required for relic abundance and dark matter core formation. 
In the off-shell Majoron case, we have found that neutrinos are not generated as a characteristic signal, unlike in the semi-annihilation case.

\section*{Acknowledgments}
This work was supported by a JSPS Grant-in-Aid for Scientific Research KAKENHI Grant No. JP20K22349.
The numerical computation in this work was carried out at the Yukawa Institute Computer Facility.

\end{document}